\documentclass[sigconf]{acmart}
\usepackage{multirow}
\usepackage{multicol}

\usepackage{subcaption}
\usepackage[flushleft]{threeparttable}
\usepackage{flushend}

\usepackage{fancyhdr}

\AtBeginDocument{%
  \providecommand\BibTeX{{%
    \normalfont B\kern-0.5em{\scshape i\kern-0.25em b}\kern-0.8em\TeX}}}

\copyrightyear{2020} 
\acmYear{2020} 
\setcopyright{acmcopyright}
\acmConference[KDD '20]{Proceedings of the 26th ACM SIGKDD Conference on Knowledge Discovery and Data Mining}{August 23--27, 2020}{Virtual Event, CA, USA}
\acmBooktitle{Proceedings of the 26th ACM SIGKDD Conference on Knowledge Discovery and Data Mining (KDD '20), August 23--27, 2020, Virtual Event, CA, USA}
\acmPrice{15.00}
\acmDOI{10.1145/3394486.3412856}
\acmISBN{978-1-4503-7998-4/20/08}

\fancypagestyle{firstpage}{%
  \lhead{}
  \chead{}
  \rhead{Published as a conference paper at KDD 2020}
}

\begin{document}
\pagestyle{fancy}
\fancyhead[L]{}
\fancyhead[R]{Published as a conference paper at KDD 2020}
\fancyhf[cf]{\thepage}

\title{Understanding the Impact of the COVID-19 Pandemic on Transportation-related Behaviors with Human Mobility Data}

\author{Jizhou Huang, Haifeng Wang, Miao Fan, An Zhuo, Yibo Sun, Ying Li}
\affiliation{%
  \institution{Baidu Inc., Beijing, China}
}
\email{{huangjizhou01, wanghaifeng, fanmiao, zhuoan, sunyibo, liying}@baidu.com}


\begin{abstract}
The constrained outbreak of COVID-19 in Mainland China has recently been regarded as a successful example of fighting this highly contagious virus. Both the short period (in about three months) of transmission and the sub-exponential increase of confirmed cases in Mainland China have proved that the Chinese authorities took effective epidemic prevention measures, such as case isolation, travel restrictions, closing recreational venues, and banning public gatherings. These measures can, of course, effectively control the spread of the COVID-19 pandemic. Meanwhile, they may dramatically change the human mobility patterns, such as the daily transportation-related behaviors of the public. To better understand the impact of COVID-19 on transportation-related behaviors and to provide more targeted anti-epidemic measures, we use the huge amount of human mobility data collected from Baidu Maps, a widely-used Web mapping service in China, to look into the detail reaction of the people there during the pandemic. To be specific, we conduct data-driven analysis on transportation-related behaviors during the pandemic from the perspectives of 1) means of transportation, 2) type of visited venues, 3) check-in time of venues, 4) preference on ``origin-destination'' distance, and 5) ``origin-transportation-destination'' patterns. For each topic, we also give our specific insights and policy-making suggestions. Given that the COVID-19 pandemic is still spreading in more than 200 countries and territories worldwide, infecting millions of people, the insights and suggestions provided here may help fight COVID-19.  
\end{abstract}



\maketitle
\thispagestyle{firstpage}  

\section{Introduction}
COVID-19 stands for coronavirus disease 2019, which is caused by the coronavirus SARS-CoV-2~\citep{wang2020genetic}. Due to its high infectivity, total confirmed cases of COVID-19 dramatically increased from approximately 300 on January 20, 2020 to more than 5.6 million worldwide on May 27, 2020.\footnote{\url{https://en.wikipedia.org/wiki/COVID-19\_pandemic}} The rapid increase of confirmed cases raised the common concern of humankind, and the COVID-19 outbreak was officially announced as a pandemic by the World Health Organization (WHO) on March 11, 2020.\footnote{\url{https://www.who.int/dg/speeches/detail/who-director-general-s-opening-remarks-at-the-media-briefing-on-covid-19---11-march-2020}}

In order to fight this highly contagious virus, the authorities of many countries have imposed different levels of containment policies. The Chinese authorities are believed to have taken effective epidemic prevention measures~\citep{Maiereabb4557,Kraemereabb4218}, such as case isolation, travel restrictions~\citep{Chinazzieaba9757}, closing recreational venues, and banning public gatherings. These anti-epidemic measures have been proven to shorten the period of transmission~\citep{Ferrettieabb6936} and result in the sub-exponential growth of confirmed cases~\citep{Maiereabb4557}, which demonstrates that the majority of them are effective to control the spread of the COVID-19.

While these strategies have effectively dealt with the critical situations of outbreaks, the combination of the pandemic and mobility controls has drastically affected the general public. For example, a recent study on user behaviors during the pandemic showed that these imposed measures could lead to significant changes in collective responses of the public~\citep{xiong2020understanding}. Besides that, \citet{huang2020quantifying} further demonstrated that the imposed mobility restrictions tend to bring a significant impact on the national economy. 

\begin{figure*}[!htp]
\centering
\begin{subfigure}{.485\textwidth}
  \centering
  \includegraphics[width=1.0\linewidth,trim={0.7cm 0.7cm 0.7cm 0.1cm},clip]{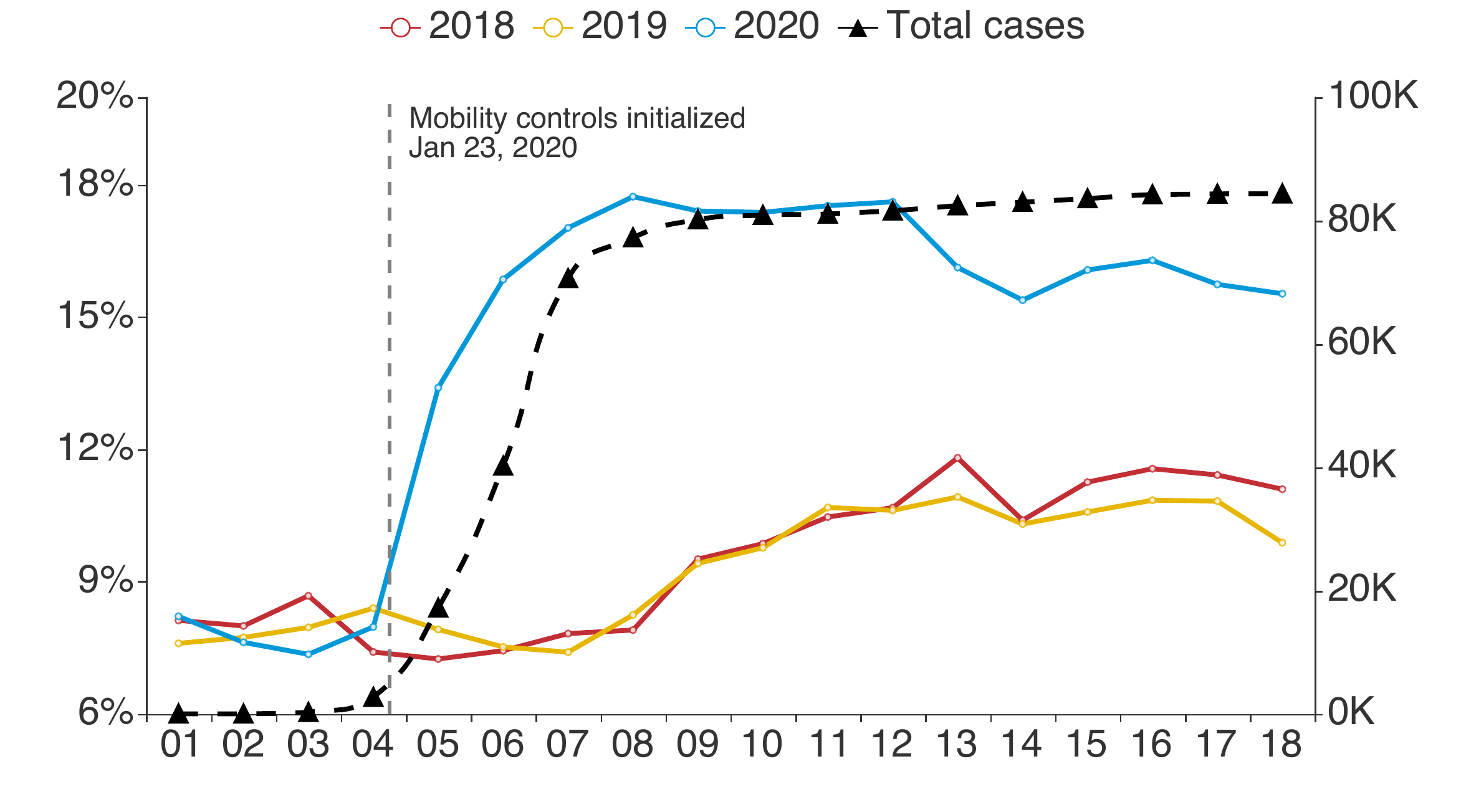}  
  \caption{By bicycle}
  \label{fig:sub-first}
\end{subfigure}
\hfill%
\begin{subfigure}{.485\textwidth}
  \centering
  \includegraphics[width=1.0\linewidth,trim={0.7cm 0.7cm 0.7cm 0.1cm},clip]{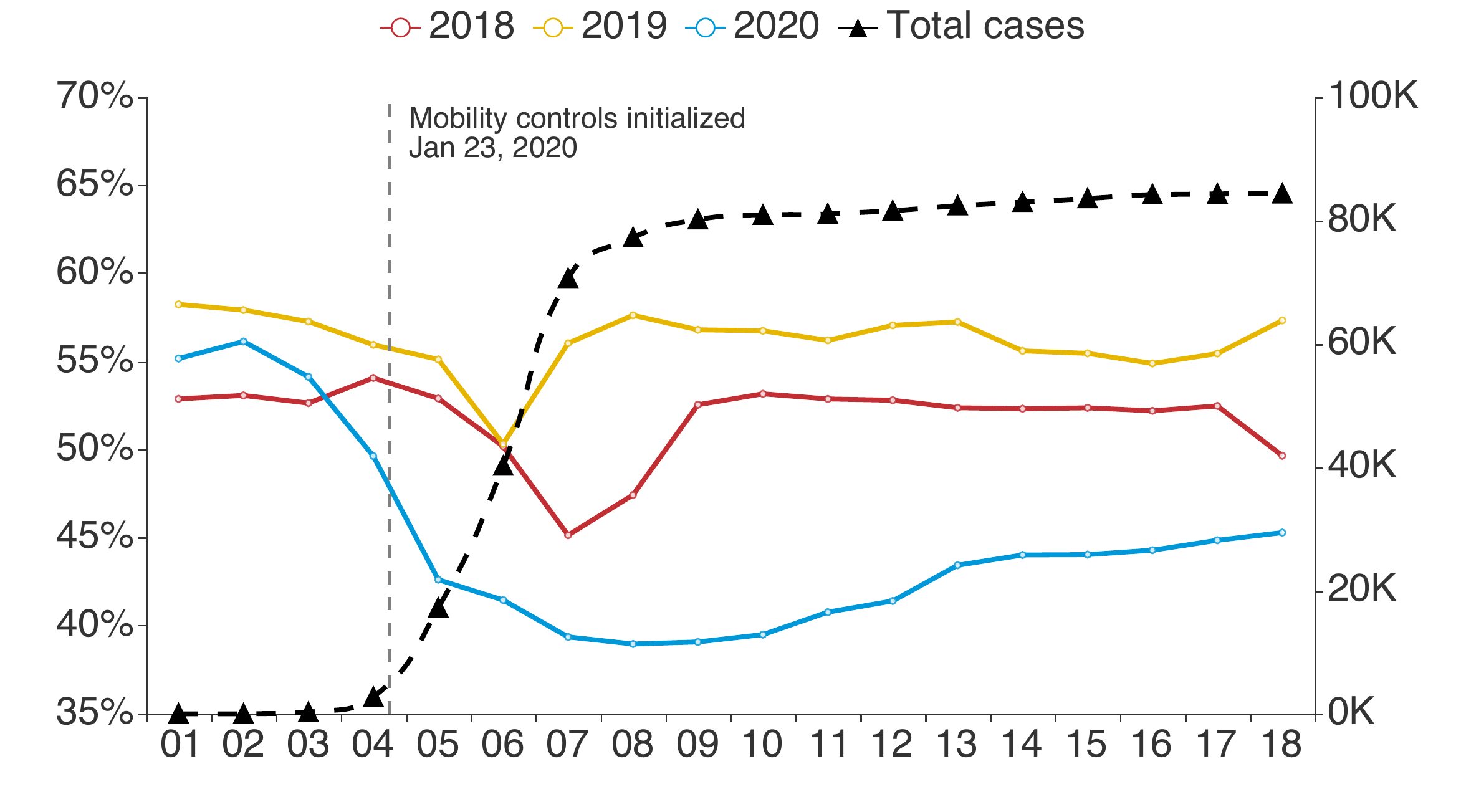}  
  \caption{By public transit}
  \label{fig:sub-second}
\end{subfigure}
\vspace{-3mm}
\caption{The weekly normalized frequency of the users in Mainland China who used Baidu Maps to navigate to their desired venues by bicycle and public transit. This figure simultaneously illustrates the weekly frequency from the 1st week to the 18th week in the years of 2018, 2019, and 2020. In addition, we plot the curve of the number of confirmed cases to find out the relationship between the severity of the COVID-19 pandemic and the means of transportation preferred by the public. }
\label{fig:week-covid-data}
\end{figure*}

We consider that the containment policy is a double-edged sword. On the one hand, the restrictions on human mobility can constrain the outbreak of the COVID-19 pandemic. On the other hand, various impact of the combination of COVID-19 and the mobility controls emerges during the pandemic. We believe a direct impact of this pandemic is on transportation-related behaviors of the public. Yet, to the best of our knowledge, the transportation-related impact has not been fully explored and better understood. Therefore, the far-reaching significance of understanding the impact of COVID-19 on transportation-related behaviors is to provide more specific policy-making suggestions on anti-epidemic measures. For example, if we could figure out which type of venue is still visited frequently by the public during the pandemic, the authorities can thus put in more effort to avoid transmitting potentially new infections in these venues.

Here comes the question that ``how do we get to know the transportation-related behaviors of the public?''. We believe that the human mobility data collected from Web mapping services could help us conduct data-driven analysis on transportation-related behaviors during the pandemic.
Figure~\ref{fig:week-covid-data} illustrates the weekly normalized frequency of the users in Mainland China who leveraged Baidu Maps, a widely-used Web mapping service in China, to navigate to their desired venues by bicycle and public transit in the first 18 weeks of the years 2018, 2019, and 2020. We also plot the curve of the number of confirmed cases to explore the relationship between the severity of the COVID-19 pandemic and the means of transportation preferred by the public. We can see from the figure that both means of transportation in the years of 2018 and 2019 is relatively consistent with strongly positive correlations (by bicycle: correlation $=96.48\%$ with p-value $= 2.53 \times 10^{-7} \leq 10^{-4}$ and by public transit: correlation $=86.19\%$ with p-value $=6.15 \times 10 ^{-6} \leq  10^{-4}$). However, as the COVID-19 began to outbreak from the 5th week of the year 2020, the correlation on the two means of transportation between the year 2019 and 2020 dramatically change (by bicycle: correlation $=54.22\%$ with p-value $=5.27 \times 10 ^{-5} \leq 10^{-4}$ and by public transit: correlation $=-5.64\%$ with p-value $=3.32 \times 10 ^ {-5} \leq 10 ^ {-4} $). Moreover, we find a positive correlation $=24.79\%$ with p-value $=3.10 \times 10 ^{-8} \leq 10 ^ {-4}$ between the number of confirmed cases and the weekly normalized frequency of transportation by bicycle, as well as a negative correlation $=-56.54\%$ with p-value $=1.36 \times 10 ^ {-5} \leq 10 ^ {-4}$ between the number of confirmed cases and the weekly normalized frequency of transportation by public transit. Both statistical results indicate that the pandemic has an impact on the two kinds of transportation-related behaviors of the public. Given that Mainland China has just passed the outbreak period of COVID-19, those initial findings from Figure~\ref{fig:week-covid-data} further drive us to look into more details on transportation-related reactions of the people there during the pandemic.

To quantify transportation-related reactions of the people in Mainland China, we decided to use the huge amount of human mobility data collected from Baidu Maps, which is one of the largest Web mapping applications with over 340 million monthly active users worldwide by the end of December 2016\footnote{\url{http://ir.baidu.com/static-files/e249a0f8-082a-4f8a-b60d-7417fa2f8e7e}}. We believe that the statistical results from the vast volumes of human mobility data are significant. The data sample is a tuple, i.e., (origin, means of transportation, destination, departure time, arrival time), which is composed of the classical OTD (origin-transportation-destination)~\citep{pitombeira2020dynamic} information as well as the time. We conduct data-driven analysis on transportation-related behaviors during the pandemic based on this data. To be specific, the transportation-related behaviors concerned in this paper include: 1) means of transportation, 2) type of visited venues, 3) check-in time of venues, 4) preference on ``origin-destination'' distance, and 5) ``origin-transportation-destination'' patterns.
These topics have covered almost all aspects of transportation-related behaviors from the perspectives of \textbf{T} (transportation), \textbf{OD} (origin \& destination venues) , and \textbf{OTD} (origin-transportation-destination triplets). According to the abnormal behaviors of the public during the pandemic, we also give our specific insights and policy-making suggestions for each perspective as follows: 
\begin{itemize}
    \item \textbf{T} (transportation): Two main insights and suggestions derive from the statistics. (1) To ride on a bicycle is strongly encouraged, and more bicycle lanes could be temporarily expanded. (2) The authorities may coordinate the working hours of various companies, and suggest the public to travel in different peaks. 
    \item \textbf{OD} (origin \& destination venues): The observations support four basic insights and suggestions. (1) For residential areas as well as hospitals \& pharmacies, the authorities need to invest more effort to limit transmission to susceptible individuals. (2) After the pandemic is over, more economic support policies should be formulated for restaurants \& beverages, parks \& scenic spots, educational institutes, and hotels. (3) Wednesday and Thursday deserve more attention, as the people in Mainland China prefer going outdoors on these days during the pandemic rather than at weekends. (4) The containment policy imposed by the Chinese authorities has been proved to effectively control the spread of COVID-19~\citep{Maiereabb4557,Kraemereabb4218}. This is also evidenced by the declined proportion of inter-city navigation requests in March and April in 2020.
    \item \textbf{OTD} (origin-transportation-destination triplets): Three insights and suggestions arise from the qualitative data analysis. (1) The most dramatic change of people's travel patterns (i.e., the top-1 OTD pattern) during the pandemic is that the start and end points of navigation by walk and private vehicle, are replaced by residential areas from transport facilities. This phenomenon indicates that people prefer zero-touch transportation to and from their places of residence. (2) Given the top-5 OTD patterns, the COVID-19 pandemic causes markets and workplaces to be the preferred destinations rather than hotels and educational institutes when people decide to go outside. We consider this phenomenon may be partly due to the containment policy, which stimulates the need to stockpile daily necessities and resume work. Therefore, the authorities need to invest more effort in such venues to avoid transmitting potentially new infections. (3) The increased proportion of taking public transit to hospitals \& pharmacies should draw much attention from the authorities, as this OTD pattern tends to exacerbate the risk of infectious disease transmission. Hence, we propose to strengthen the prevention measures at transport facilities. 
\end{itemize}

As the COVID-19 pandemic is still spreading in more than 200 countries and territories worldwide, infecting more than 5.6 million people around the world as of May 27, 2020, these insights and suggestions may help people better fight this disease. Moreover, we hope that the analysis and insights could further inspire the following work on predicting the prevalence of other similar infectious diseases with human mobility data.

\section{Impact of COVID-19 on Transportation-related Behaviors}
In this section, we use the huge amount of human mobility data collected from Baidu Maps to quantify transportation-related behaviors of the people in Mainland China. Since the outbreak of COVID-19 in Mainland China started from January 2020, we mainly collect the human mobility data from January 2020 to April 2020. For comparison purposes, we also need to collect more data in the same period of the years 2018 and 2019.

The transportation-related behaviors are ``recorded and encoded'' into the canonical OTD (origin-transportation-destination) information~\citep{pitombeira2020dynamic} as part of human mobility data. Hence, we will look into the details of the transportation-related behaviors of the people in Mainland China from the perspectives of \textbf{T} (transportation), \textbf{OD} (origin \& destination venues), and \textbf{OTD} (origin-transportation-destination triplets).
To be specific, we will conduct data-driven analysis on means of transportation (\S \ref{subsec:means_of_transportation}), type of visited venues (\S \ref{subsec:type-of-visited-venues}), check-in time of venues (\S\ref{subsec:check-in_time_of_venues}), preference on ``origin-destination'' distance (\S \ref{subsec:preference-od-distance}), and  ``origin-transportation-destination'' patterns (\S \ref{subsec:otd-patterns}).

\subsection{Means of Transportation}
\label{subsec:means_of_transportation}

\begin{figure}
\centering
\includegraphics[width=0.478\textwidth,trim={0.9cm 0.2cm 1.2cm 0.1cm},clip]{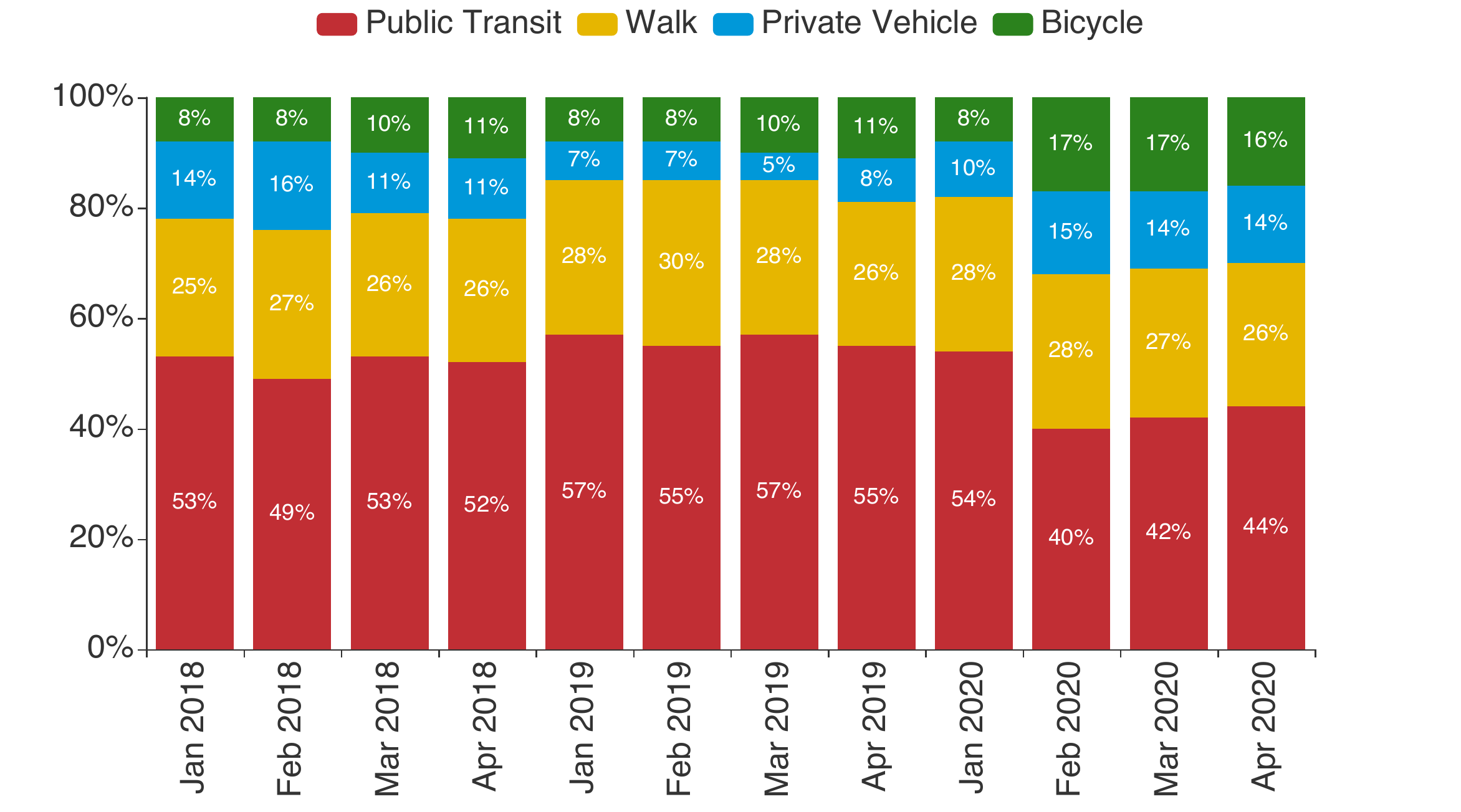}
\caption{The normalized frequency of taking various means of transportation recorded by Baidu Maps in Mainland China. We illustrate the monthly data from January to April in the years of 2018, 2019, and 2020, respectively. Each bar represents the proportions of different means of transportation by month.}
\label{fig:means-of-transportation}
\end{figure}

\subsubsection{Data Processing \& Illustration} 
We collect the navigation data recorded by Baidu Maps in Mainland China from January to April in the years of 2018, 2019, and 2020, respectively. The monthly frequency of the people there who used Baidu Maps to go out is calculated and categorized by various means of transportation.
The statistical results are illustrated by Figure~\ref{fig:means-of-transportation}, where each bar represents the proportions of different means of transportation by month. The means of transportation include using public transit, taking a private vehicle, riding on a bicycle, and walking.

\subsubsection{Data Analysis \& Insights}
From Figure~\ref{fig:means-of-transportation}, we can see that the proportions of the means of transportation from January to April in the years of 2018 and 2019 are highly stable and consistent. Without the impact of COVID-19 pandemic during these periods, the people in Mainland China prefer taking public transit as $54\%$ navigation requests on average\footnote{This value is the average of eight months' proportions of public transit from January to April in the years of 2018 and 2019.} at Baidu Maps chose to take this means of transportation to reach the destinations. However, the proportion of taking public transit declines to $45\%$ in the first four months of the year 2020. What's worse, along with the most severe outbreak in February 2020, the proportion of taking public transit dramatically declines to $40\%$. On the contrary, the proportions of taking a private vehicle and riding on a bicycle significantly increase during the outbreak of COVID-19. Compared with the statistics at the same period in 2018 and 2019, the proportion of taking a private vehicle increases by $3.38\%$ and the proportion of riding a bicycle increases by $5.25\%$. Both changes indicate that more people prefer taking private transportation during the pandemic, which can minimize human contact. 

\subsubsection{Policy-making Suggestions}
According to the insights on the data analysis on means of transportation, we find out that the people in Mainland China prefer taking private transportation such as taking a private vehicle and riding on a bicycle to minimize human contact rather than taking public transit. This inevitably leads to a great increment in traffic pressure. 
To ease the traffic pressure, our policy-making suggestions are as follows:
\begin{itemize}
\item Given the possibility of future outbreaks of COVID-19 or other emerging infectious diseases, the authorities could encourage more people to travel by bicycle and consider temporarily expanding the dedicated bicycle lanes.
\item The authorities may coordinate the working hours of various companies, and suggest the public to travel in different peaks.
\end{itemize}

\subsection{Type of Visited Venues}
\label{subsec:type-of-visited-venues}

\subsubsection{Data Processing \& Illustration}
We collect the check-in data recorded by Baidu Maps in Mainland China from January to April in the years of 2018, 2019, and 2020, respectively. The monthly frequency of the people there who used Baidu Maps to visit venues is calculated and categorized by various types of venues. Figure~\ref{fig:type-of-visited-venues} illustrates the statistical results via a bar chart, where each bar represents the proportions of the top-10 hottest types of visited venues over the three years. The top-10 hottest types of visited venues include residential areas, transport facilities, shopping \& markets, restaurants \& beverages, parks \& scenic spots, educational institutes, hotels, workplaces, hospitals \& pharmacies, and government agencies.

\subsubsection{Data Analysis \& Insights}
\begin{figure}[!thp]
\centering
\includegraphics[width=0.478\textwidth,trim={0.6cm 0.2cm 1.2cm 0.1cm},clip]{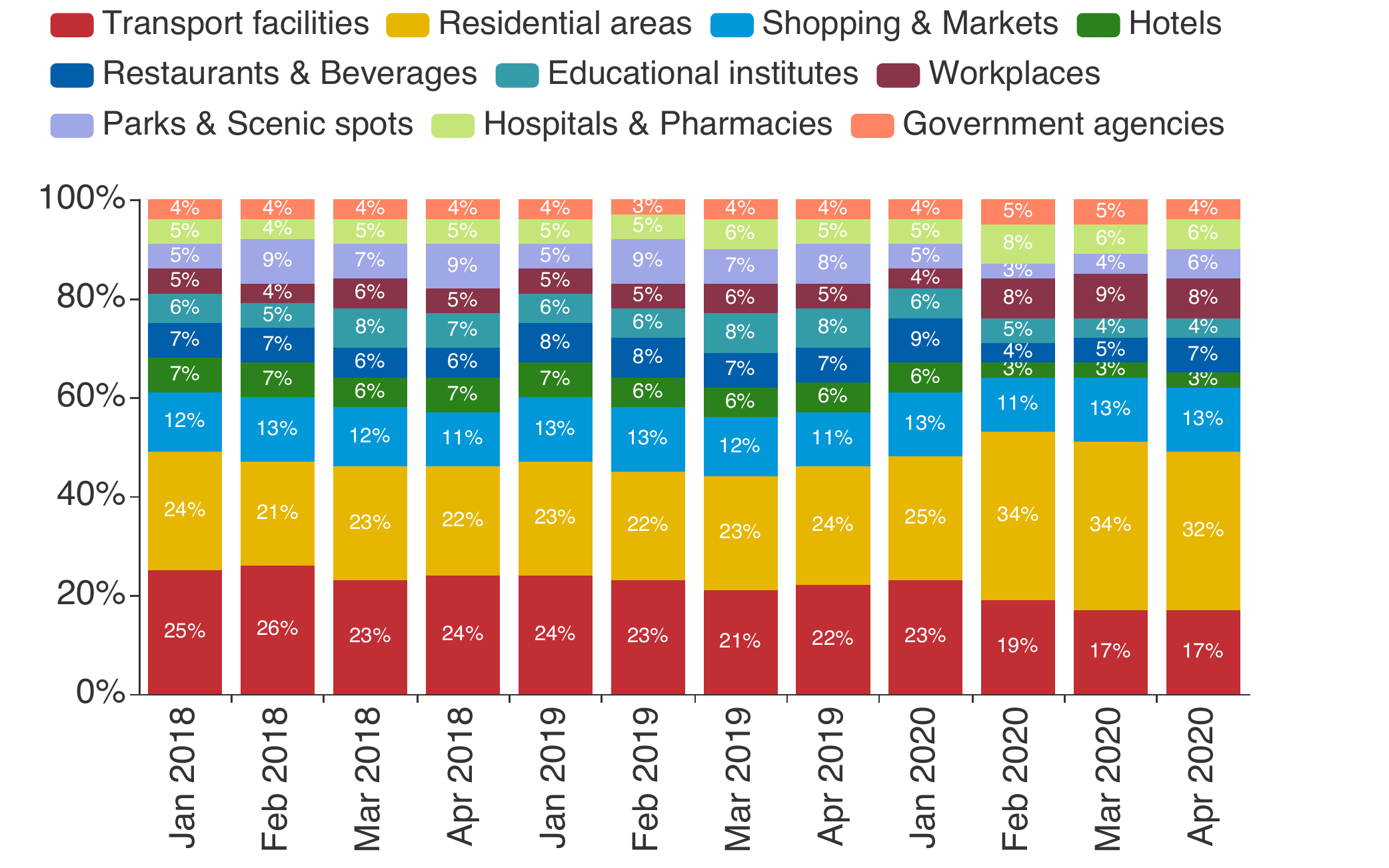}
\caption{The normalized frequency of different types of visited venues recorded by Baidu Maps in Mainland China. We use the monthly data from January to April in the years of 2018, 2019, and 2020, respectively, and illustrate the top-10 hottest types of visited venues over the three years. Each bar represents the proportions of an individual type.}
\label{fig:type-of-visited-venues}
\end{figure}

We can see from Figure~\ref{fig:type-of-visited-venues} that the distribution of visiting frequency of the top-10 hottest types of venues are highly stable and consistent from January to April in the years of 2018 and 2019. Without the impact of the COVID-19 pandemic, the proportions of visiting residential areas and transport facilities are stable at $22.75\%$ and $23.50\%$ on average, respectively. When the COVID-19 outbreaks, the proportion of visiting residential areas greatly increases to $31.25\%$, and the proportion of visiting transport facilities rapidly declines to $19.00\%$. Both alterations indicate that more people might prefer staying indoors as well as limit the frequency of going outdoors, which can minimize both human contact and the risk of infection during the pandemic. The proportion of hospitals \& pharmacies slightly increases by $1.25\%$ due to the COVID-19 pandemic. As for other types such as restaurants \& beverages, parks \& scenic spots, educational institutes, and hotels, however, the visiting frequency of them in 2020 shrinks compared with those in the same period of 2018 and 2019. 

\subsubsection{Policy-making Suggestions}
According to the insights on the data analysis on types of visited venues, we find out that the people in Mainland China prefer staying indoors or visiting hospitals \& pharmacies during the pandemic. This phenomenon may suggest the authorities to make policies as follows:
\begin{itemize}
\item For residential areas as well as hospitals \& pharmacies, the authorities need to invest more effort in these venues to limit transmission to susceptible individuals and minimize the risk of cross-infection.
\item After the pandemic is over, more economic support policies should be formulated for restaurants \& beverages, parks \& scenic spots, educational institutes, and hotels.
\end{itemize}

\subsection{Check-in Time of Venues}
\label{subsec:check-in_time_of_venues}

\begin{figure*}[!htp]
\centering
\begin{subfigure}{.325\textwidth}
  \centering
  \includegraphics[width=1.0\linewidth,trim={1.0cm 0.8cm 2.3cm 1.2cm},clip]{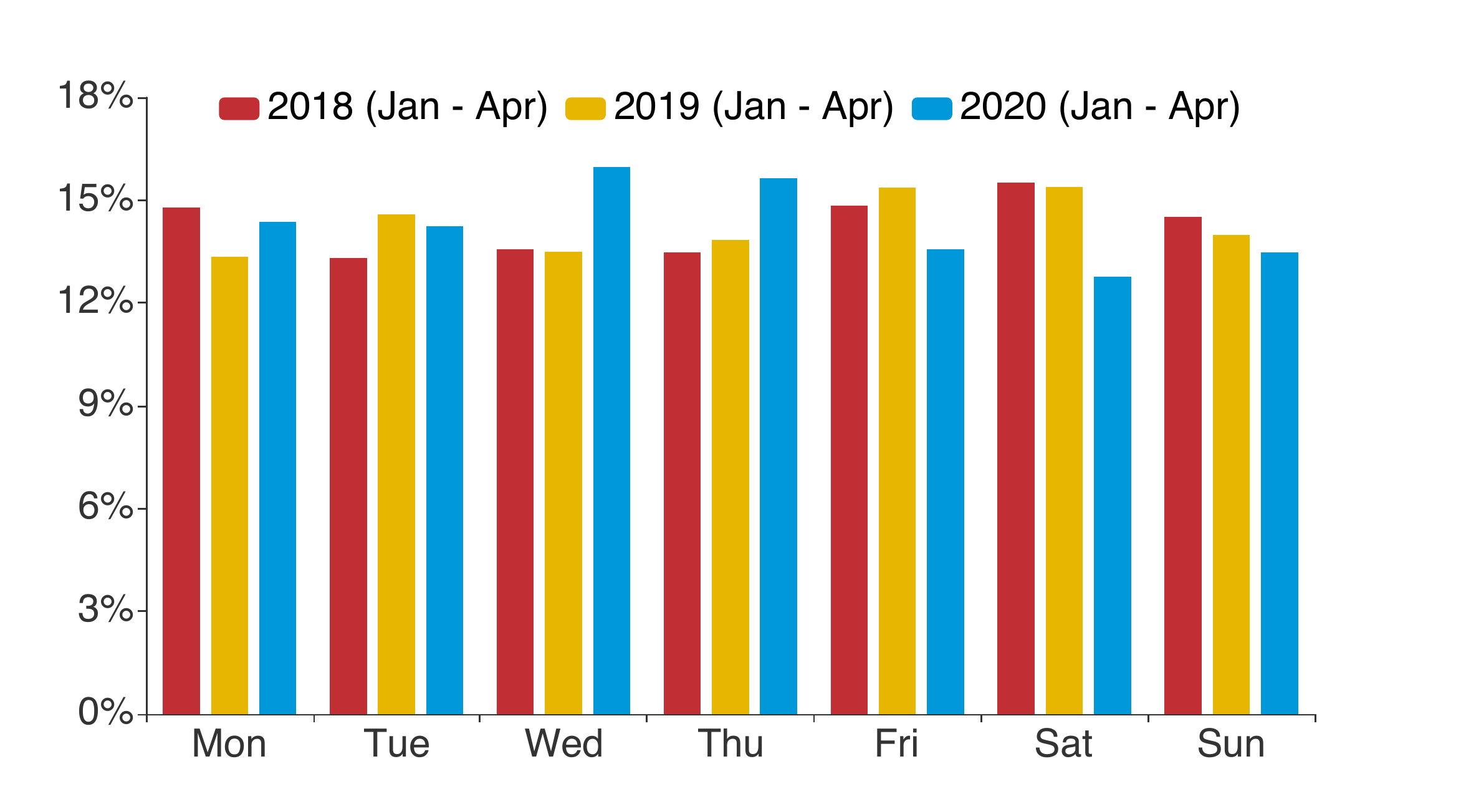}  
  \caption{All day}
  \label{fig:checkin-all}
\end{subfigure}
\hfill
\begin{subfigure}{.325\textwidth}
  \centering
  \includegraphics[width=1.0\linewidth,trim={1.0cm 0.8cm 2.3cm 1.2cm},clip]{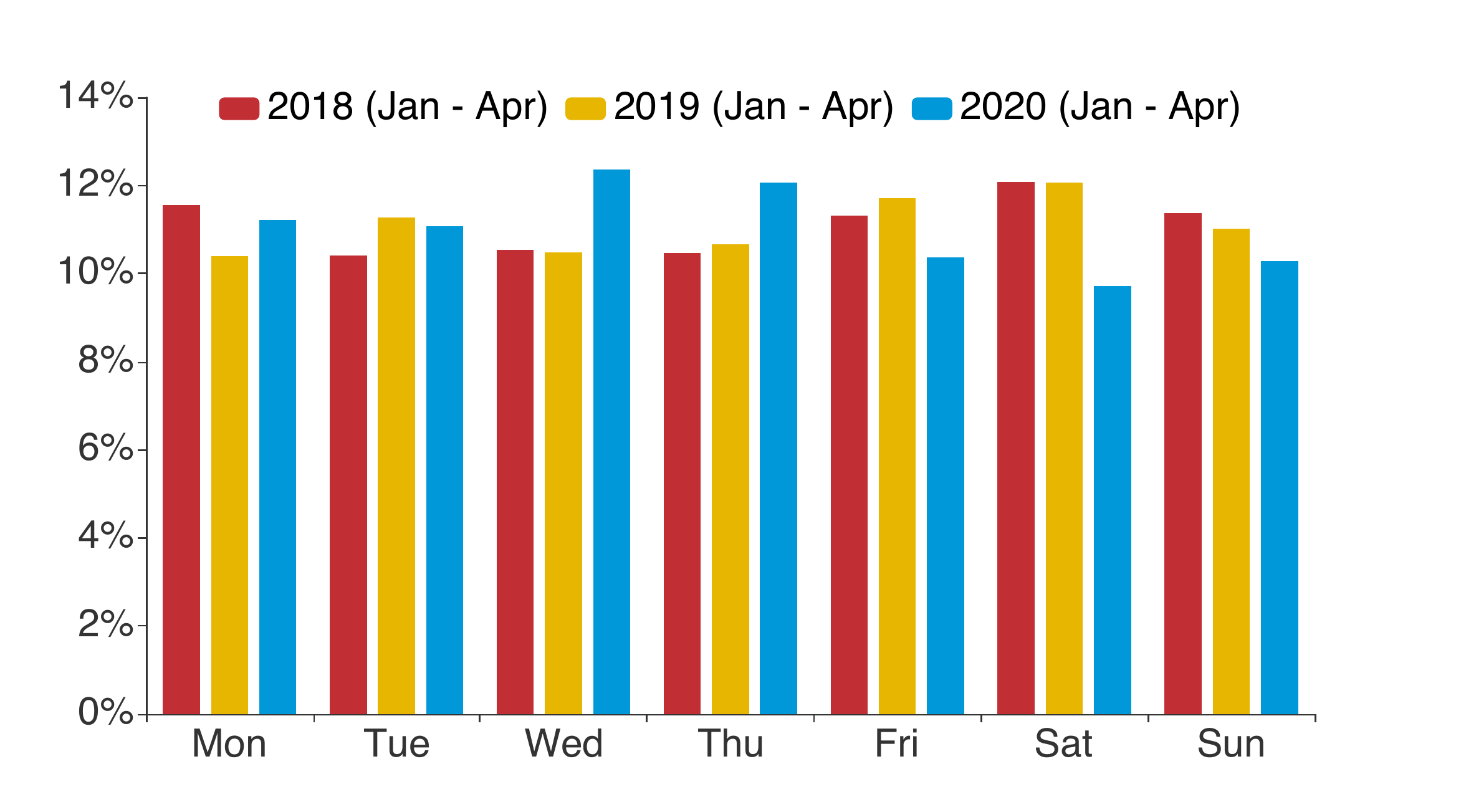}  
  \caption{Daylight hours (from 6:00 to 18:00) }
  \label{fig:checkin-day}
\end{subfigure}
\hfill%
\begin{subfigure}{.325\textwidth}
  \centering
  \includegraphics[width=1.0\linewidth,trim={1.0cm 0.8cm 2.3cm 1.2cm},clip]{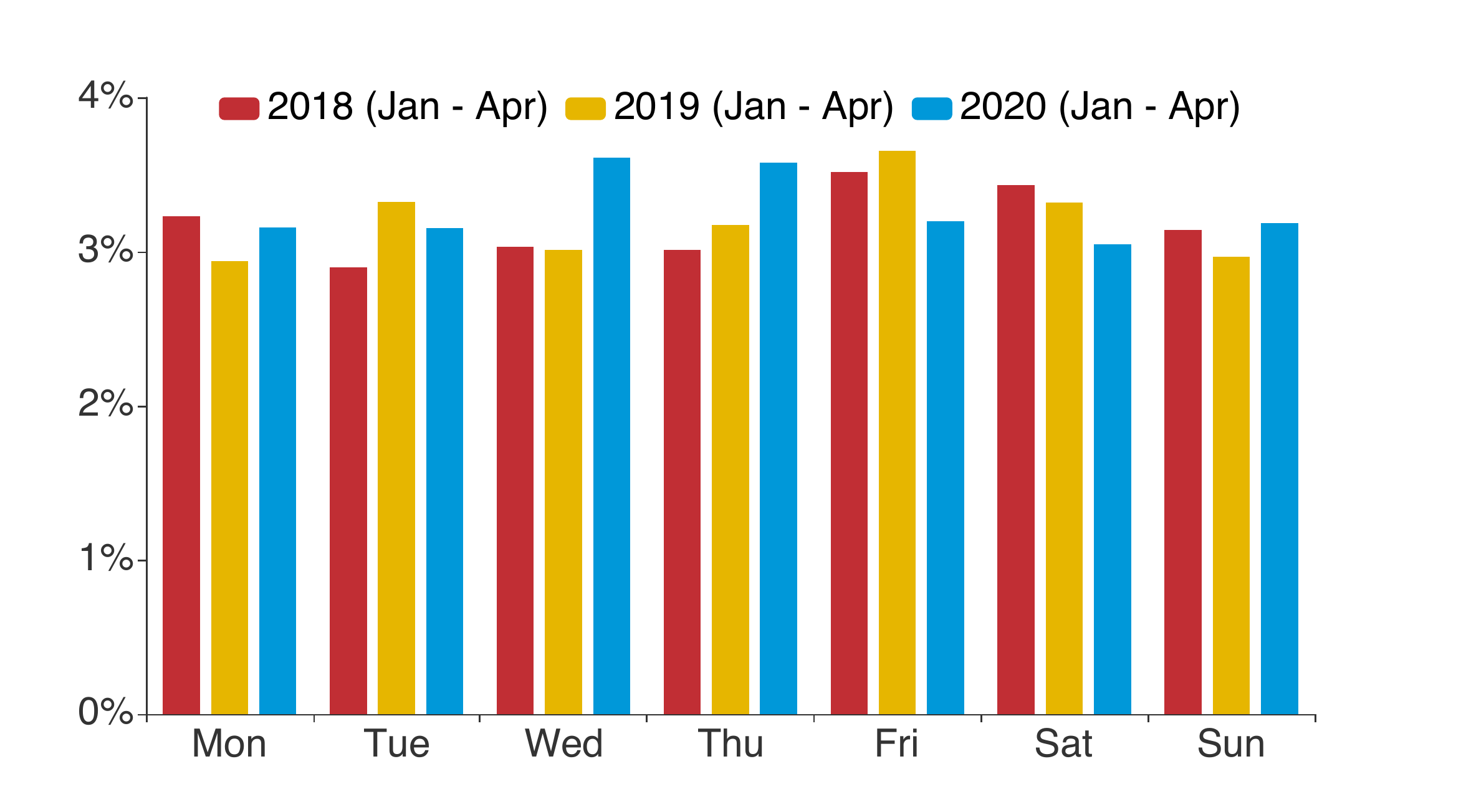}  
  \caption{Night hours (from 18:00 to 6:00 next day)}
  \label{fig:checkin-night}
\end{subfigure}
\caption{The histograms of the normalized frequency of the check-in time of venues within a week recorded by Baidu Maps in Mainland China. We leverage the data from January to April in the years of 2018, 2019, and 2020, respectively. The check-in time of venues is categorized into 7 classes (from the day of Monday to Sunday night), indicating the daylight hours (from 6:00 to 18:00) and night hours (from 18:00 to 6:00 next day) of the seven days within a week.}
\label{fig:week-check-in}
\end{figure*}

\subsubsection{Data Processing \& Illustration}
We accumulate the check-in time recorded by Baidu Maps in Mainland China from January to April in the years of 2018, 2019, and 2020, respectively. As illustrated by Figure~\ref{fig:checkin-all}, the check-in time of venues is categorized into 7 classes, indicating the seven days within a week. To be specific, Figure~\ref{fig:checkin-day} displays the histogram of the normalized frequency of the check-in behaviors in the daylight hours from January to April in the years of 2018, 2019, and 2020. Figure~\ref{fig:checkin-night} shows another histogram of the normalized frequency of the check-in behaviors in the night hours from January to April in the years of 2018, 2019, and 2020.

\subsubsection{Data Analysis \& Insights}
To check out the consistency of the check-in time without the impact of COVID-19, we calculate the correlation between the histograms of 2018 and 2019. Given the three sub-figures in Figure~\ref{fig:week-check-in}, we can obtain three positive correlations (i.e., $47.68\%$, $50.93\%$, and $50.53\%$) between the histograms of 2018 and of 2019 regardless of all day, daylight hours, and night hours. The check-in time in 2018 and 2019 have the same characteristics that people prefer going outside at weekends without the panic of the contiguous disease. However, we gain strongly negative correlations (i.e., $-87.01\%$, $-93.32\%$, and $-42.56\%$) between the histogram of 2020 and the averaged histogram of 2018 and 2019, indicating that the outbreak of COVID-19 have tremendously changed the way people visit and check in venues.

\begin{figure*}[!ht]
  \centering
  \begin{subfigure}{.325\textwidth}
    \centering
    \includegraphics[width=1.0\linewidth,trim={0.5cm 0.3cm 2.3cm 0.1cm},clip]{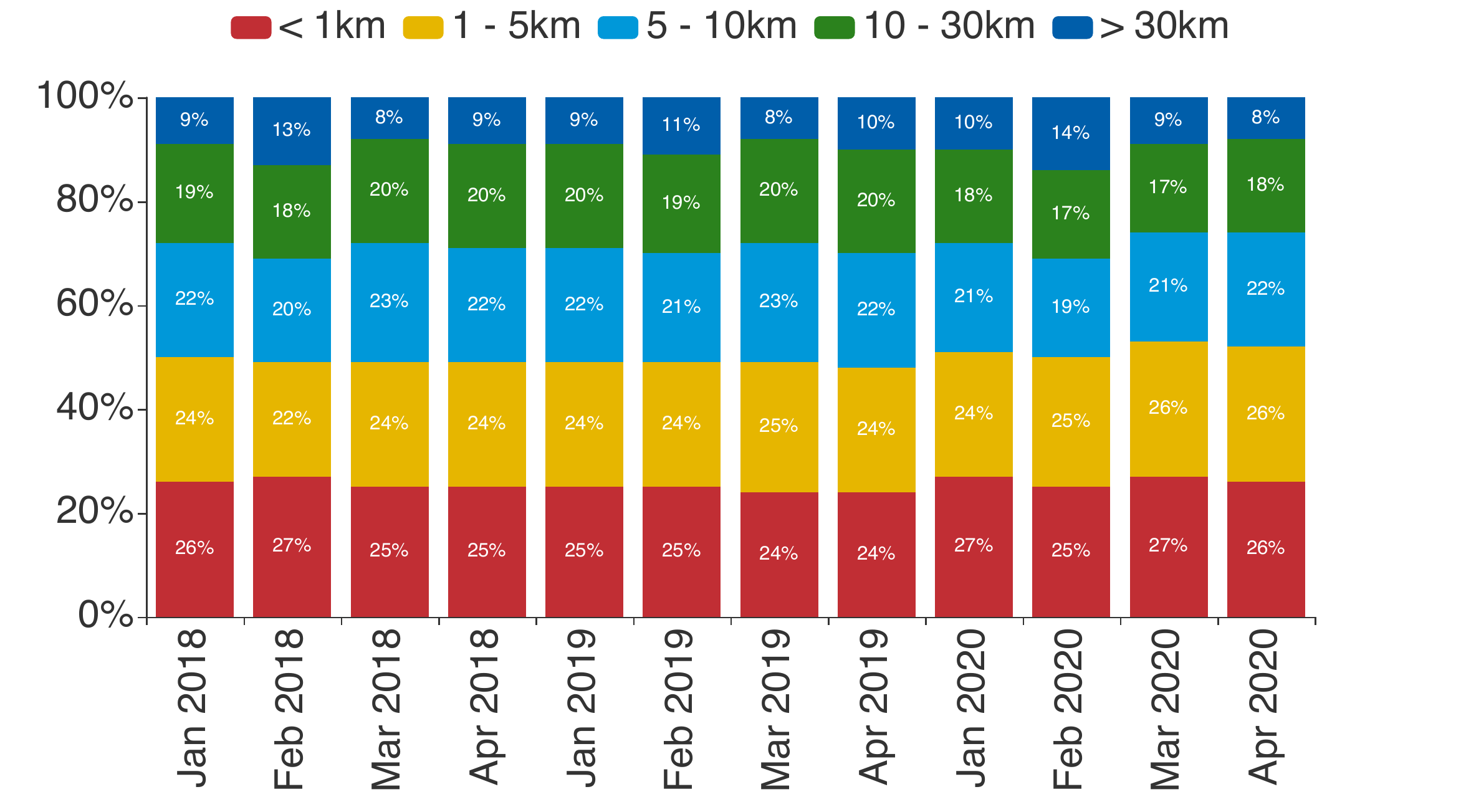}  
    \caption{The distribution of the OD distance discretized by different buckets. }
    \label{fig:od-dis-all}
  \end{subfigure}
  \hfill%
  \begin{subfigure}{.325\textwidth}
    \centering
    \includegraphics[width=1.0\linewidth,trim={0.5cm 0.3cm 2.3cm 0.1cm},clip]{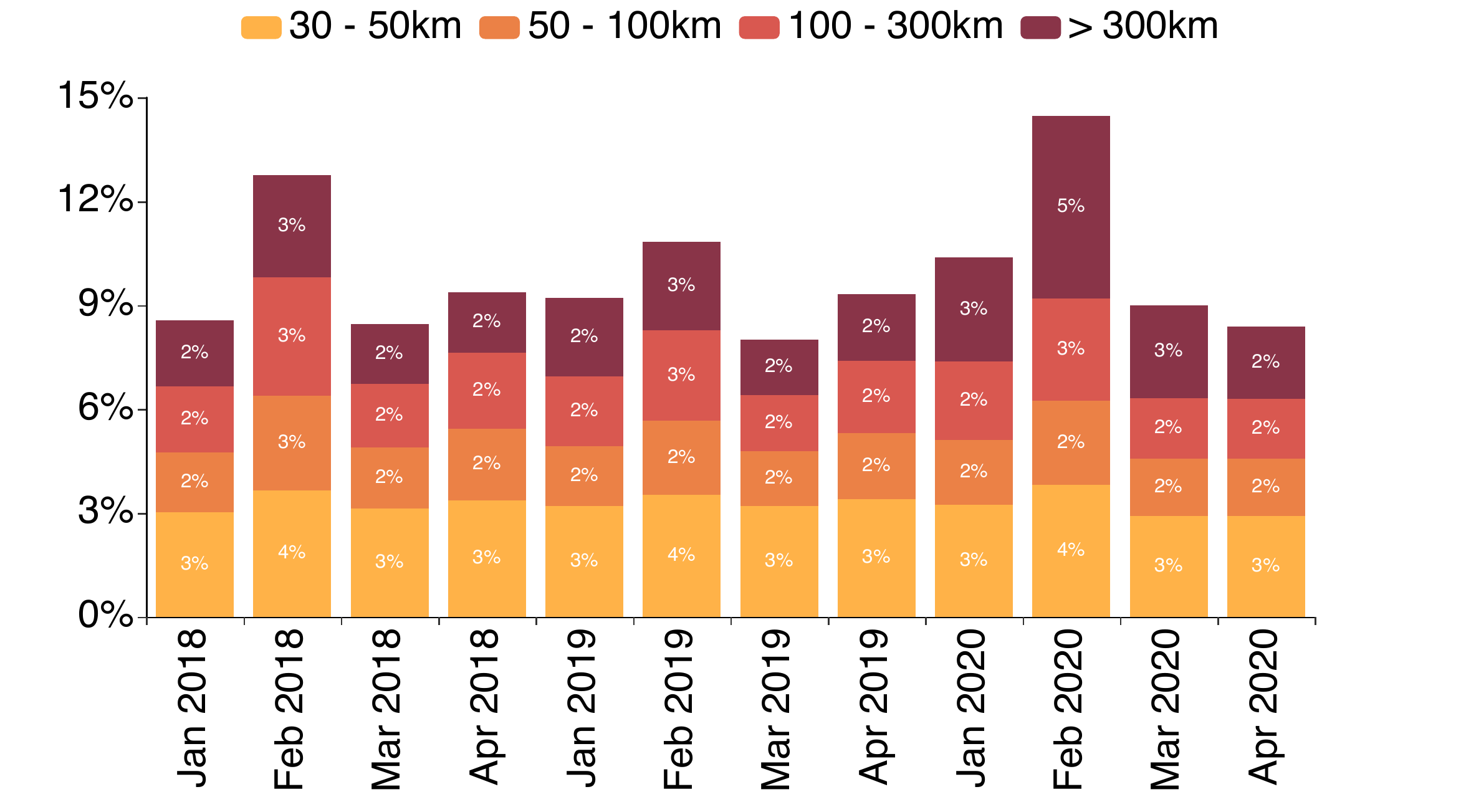}  
    \caption{The specific distribution of the OD distance $> 30km$ discretized by different buckets. }
    \label{fig:od-dis-300}
  \end{subfigure}
  \hfill%
  \begin{subfigure}{.325\textwidth}
    \centering
    \includegraphics[width=1.0\linewidth,trim={0.5cm 0.3cm 2.3cm 0.1cm},clip]{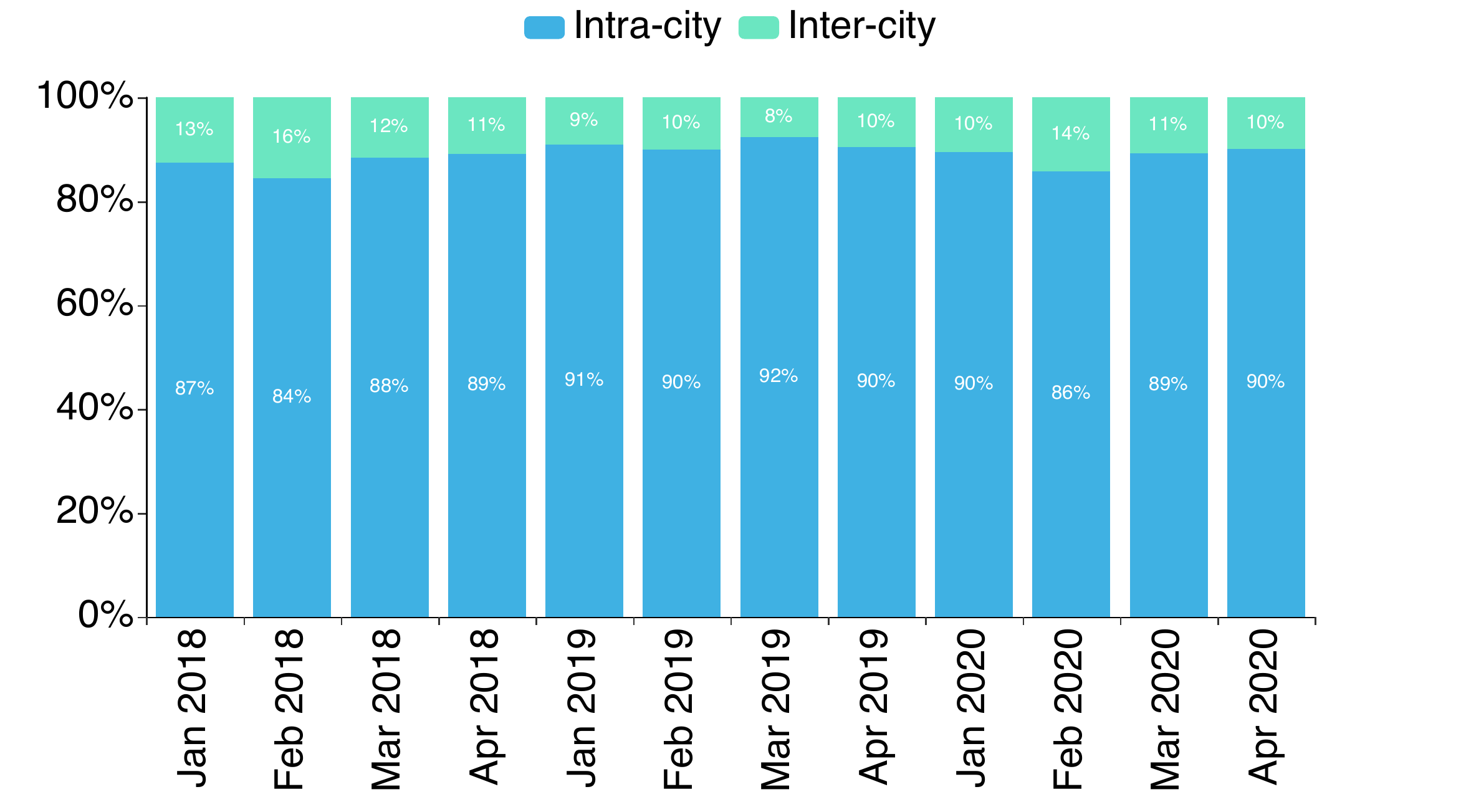}  
    \caption{The distribution of normalized frequency of intra- and inter- city traveling behaviors. }
    \label{fig:od-inter-intra}
  \end{subfigure}

  \caption{The histograms related to the ``origin-destination'' distance recorded by Baidu Maps in Mainland China. We use the human mobility data collected from January to April in the years of 2018, 2019, and 2020, respectively.}
  \label{fig:preference-od-distance}

\end{figure*}

\subsubsection{Policy-making Suggestions}
According to the insights on the data analysis on check-in time of venues, we surprisingly find out that the people in Mainland China prefer going outdoors on Wednesday and Thursday during the pandemic rather than at weekends. This observation suggests the authorities to pay more attention on both the daylight and night hours of Wednesday and Thursday rather than weekends.

\subsection{Preference on ``Origin-Destination'' Distance}
\label{subsec:preference-od-distance}

\subsubsection{Data Processing \& Illustration}
Figure~\ref{fig:preference-od-distance} illustrates a series of sub-figures related to the ``origin-destination'' distance of navigation recorded by Baidu Maps in Mainland China. We first categorize the OD distance into different buckets for discretization, and then compute the proportions of each bucket by month. The result is shown by Figure~\ref{fig:od-dis-all}. 
As we observe an increment of the long distance preference (i.e., $>30km$) in February 2020, the worst period of the pandemic in China, we decide to look into the specific distribution of the OD distance $>30km$ in Figure~\ref{fig:od-dis-300}. The bucket of distance $>300km$ in Figure~\ref{fig:od-dis-300} is amplified in February 2020, which is probably caused by the behaviors of inter-city navigation. In order to make quantitative analysis on this, we further present Figure~\ref{fig:od-inter-intra} to figure out the distribution of normalized frequency of intra- and inter- city traveling behaviors.

\subsubsection{Data Analysis \& Insights}
Figure~\ref{fig:od-dis-all} shows an abnormal proportion of the OD distance in February 2020 which is the worst period of the pandemic in China. To be specific, the proportion of the long distance preference (i.e., $>30km$) in February 2020 increases by $2\%$ in comparison with that averaged over those in February 2018 and February 2019. Furthermore, Figure~\ref{fig:od-dis-300} amplifies the distance $>30km$. From which we can see that the proportion of OD distance $>300km$ in February 2020 is significantly greater than those in February 2018 and February 2019. 

\subsubsection{Policy-making Suggestions}
The containment policy imposed by the Chinese authorities has been proved to effectively control the spread of COVID-19~\citep{Maiereabb4557}. The impact of this policy can be evidenced by the declined proportion of inter-city navigation requests, as shown by Figure~\ref{fig:od-inter-intra}, in March and April in 2020.

\subsection{``Origin-Transportation-Destination'' Patterns}
\label{subsec:otd-patterns}

\begin{table*}[!thp]
\centering
  \caption{The top-20 hottest ``origin-transportation-destination'' patterns grouped by four means of transportation, i.e., walking, riding on a bicycle, taking public transit, and taking a private vehicle. Each group of pattern is ranked in descend order based on the frequency recorded by Baidu Maps in Mainland China from January to April in the years of 2018, 2019, and 2020.}
  \vspace{-2.5mm}
  \includegraphics[width=0.99\textwidth]{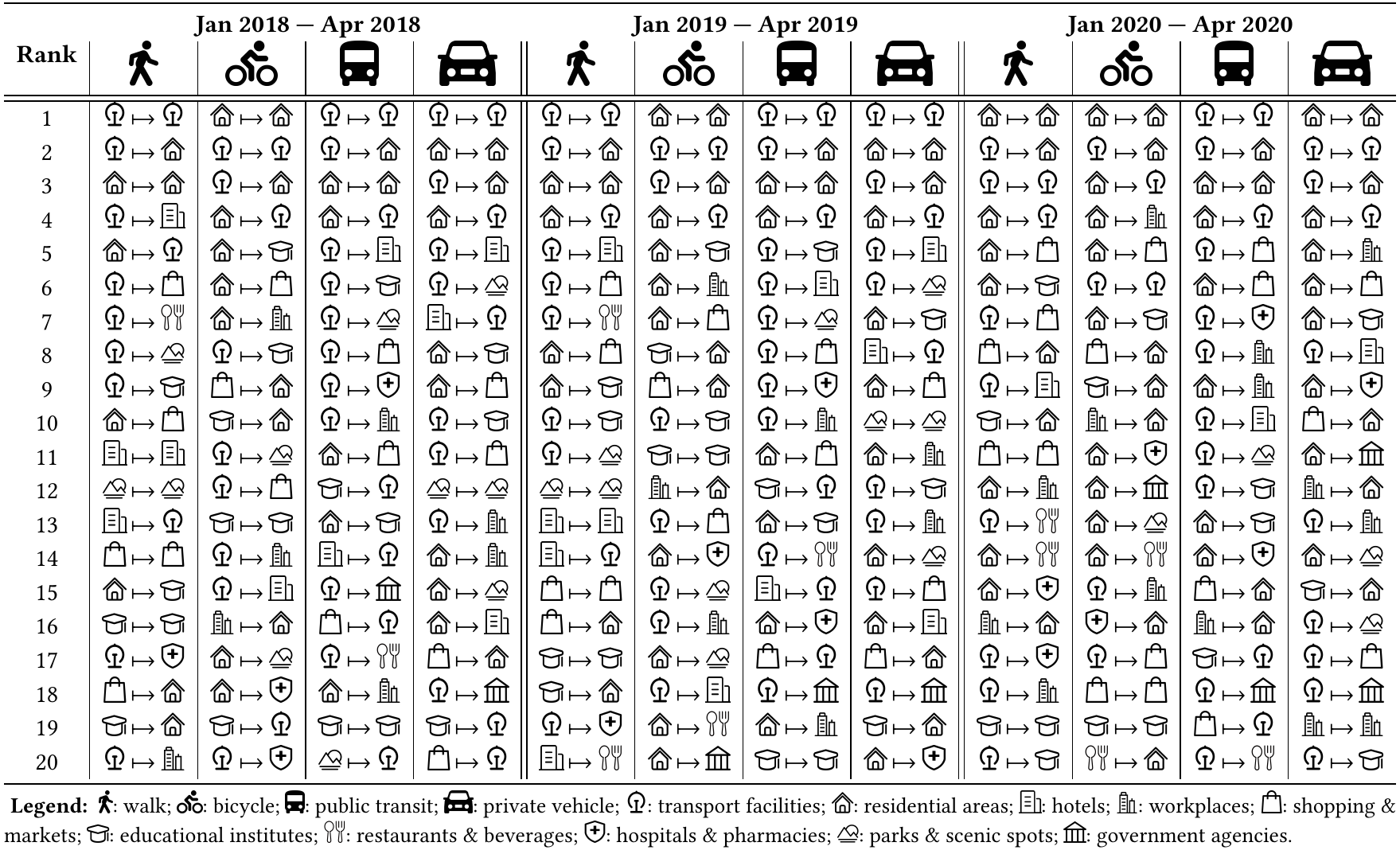}
  \label{tab:otd-top-patterns}
\end{table*}

\subsubsection{Data Processing \& Illustration}
The transportation-related behaviors are mainly ``recorded and encoded'' into the canonical OTD (origin-transportation-destination) information~\citep{pitombeira2020dynamic} as part of human mobility data. Therefore, we believe the change of OTD patterns can directly reflect the impact of the COVID-19 pandemic on transportation-related behaviors of the public. We explore the high-frequent OTD patterns from January to April in the years of 2018, 2019, and 2020, respectively, and demonstrate the top-20 hottest ``origin-transportation-destination'' patterns grouped by the four means of transportation in Table~\ref{tab:otd-top-patterns}. In addition, we use the type of venues to denote the origin and the destination instead of the specific POI (point of interest) names of them for the purpose of generalization. 

\subsubsection{Data Analysis \& Insights}
The most dramatic change of people's travel patterns (i.e., the top-1 OTD pattern) during the pandemic in the year 2020 is that the start and end points of navigation are replaced by residential areas from transport facilities. Moreover, the public prefers walking, riding on a bicycle, and taking a private vehicle rather than taking public transit. Taking public transportation means contacting a lot of people, and the general public would feel a higher risk of being infected through contact. As for walking, driving vehicles, and riding bicycles, the contact between people is more controllable, and the general public would feel that the risk of infection through contact is relatively lower. This change implies that people are willing to take zero-touch transportation to and from their places of residence to lower the risk of infection, and maybe to enhance physical fitness~\citep{doi:10.5465/AMBPP.2019.11834abstract}.

Starting from residential areas or transport facilities, the people in Mainland China would like to visit hotels and educational institutes in line with the top-5 OTD patterns in the years of 2018 and 2019. However, the destinations have entirely changed from hotels and educational institutes to workplaces and markets based on the top-5 OTD patterns during the pandemic in 2020. This phenomenon may be partly due to the containment policy, which stimulates the need to stockpile daily necessities and resume work.

In the years of 2018 and 2019, the top-10 OTD patterns cover $9$ types of destination, i.e., transport facilities, residential areas, hotels, workplace, shopping \& markets, educational institutes, restaurants \& beverages, hospitals \& pharmacies, and parks \& scenic spots. However, the diversity of the type of destination shrinks when the pandemic comes. Only $7$ types of destination are covered by the top-10 OTD patterns in the year 2020. To be specific, as destinations, restaurants \& beverages as well as parks \& scenic spots disappear from the top-10 OTD patterns in 2020 in comparison with those in the years of 2018 and 2019. Figure ~\ref{fig:type-of-visited-venues} also confirms this result. We believe that this result is mainly caused by the countermeasures carried out by the Chinese authorities~\citep{Maiereabb4557,Kraemereabb4218,Tianeabb6105}, including quarantines, travel restrictions, social distancing, digital contact tracing, suspending intra-city public transport, shutting down the recreation venues, and banning public gatherings. These interventions have effectively curbed the spread of COVID-19~\citep{Maiereabb4557}, but meanwhile they do have a significant impact on travel behavior as well as public transport supply and needs.

In addition, the rank of hospitals \& pharmacies as the destination jumps to seventh with the aid of public transit. This change needs to be given enough attention because the potential suspects are more likely to spread the disease by taking public transportation.

\subsubsection{Policy-making Suggestions}
The increased proportion of taking public transit to hospitals \& pharmacies should draw much attention from the authorities, as this OTD pattern tends to exacerbate the risk of infectious disease transmission. Therefore, we propose to strengthen the prevention measures at transport facilities. 
Moreover, the COVID-19 pandemic causes markets and workplaces to be the preferred destinations rather than hotels and educational institutes when people decide to go outside. 
To stockpile daily necessities and to resume work become the preferred choices for people in Mainland China rather than checking in hotels and visiting educational institutes. Therefore, there is a strong need to strengthen prevention measures in markets and workplaces.

\section{Emerging Studies on COVID-19}
The recent outbreak of COVID-19 has drawn much attention from many researchers around the world~\citep{sohrabi2020world,sahin20202019}. 
More than 50,000 publications mentioned the keyword ``COVID-19'' have been archived and indexed by Google Scholar in the first five months of the year 2020. In this section, we briefly review the emerging studies on COVID-19 from the perspectives of data science and epidemiology. 

\subsection{COVID-19 \& Data Science}
\citet{rao2020identification} proposed a machine learning approach Augmented ARGONet, which combines disease estimates from mechanistic models with digital traces from multiple sources via clustering and data augmentation technique, to forecast COVID-19 activity in Chinese provinces two days ahead of the current time.
\citet{liu2020machine} proposed an algorithm to identify COVID-19 cases using a mobile phone-based web survey. 
\citet{huang2020quantifying} presented a quantitative analysis to project the impact of COVID-19 on economies in Mainland China, from the perspective of mobility, with two economic indicators: New Venues Created and Volumes of Visits to Venue using the data of Baidu Maps.
\citet{xiong2020understanding} analyzed the degree of the massive population responses to the emergencies of the COVID19 pandemic in Mainland China using the real-time and historical data collected from Baidu Maps and Baidu search engine.

The world's top-tier academic journal, \textit{Science} has created a special column to publish emerging research on the COVID-19 pandemic. Several articles use data-driven analysis on COVID-19. For instance, 
\citet{Chinazzieaba9757} used a global metapopulation disease transmission model to project the impact of travel limitations on the national and international spread of the pandemic.
\citet{Ferrettieabb6936} explored the feasibility of protecting the population using isolation with contract tracing data from questionnaires and mobile phone applications.
\citet{Kraemereabb4218} used real-time mobility data from Wuhan and detailed case data, including travel history to elucidate the role of case importation in transmission in cities across China and to ascertain the impact of control measures.
\citet{Maiereabb4557} introduced a parsimonious model that captures both quarantine of symptomatic infected individuals, as well as population-wide isolation practices in response to containment policies or behavioral changes. 
\citet{Tianeabb6105} performed a quantitative analysis of the impact of China's control measures of COVID-19 using a dataset that includes case reports, human movement, and public health interventions.

\subsection{COVID-19 \& Epidemiology}
In the research field of epidemiology, \citet{surveillances2020epidemiological} reported results of a descriptive, exploratory analysis of all COVID-19 cases extracted from China's Infectious Disease Information System through February 11, 2020.
To address the challenges of robust collection of population-scale data for COVID-19, \citet{Drew1362} developed a mobile software COVID Symptom Study which encourages reporting of potential COVID-19 symptoms. They recruited about 2 million users to the COVID Symptom Study from across the United Kingdom and the United States. Based on the epidemiologic data rapidly collected by the software, mathematical modeling predicted geographical hotspots of incidence 5 to 7 days in advance of official public health reports in Wales, United Kingdom. 
To understand the transmission future of SARS-CoV-2, which is the coronavirus responsible for the current COVID-19 pandemic. \citet{Kissler860} used existing time-series data from the United States to build a deterministic model of multiyear interactions between existing coronaviruses, and used this to project the potential pandemic dynamics and pressures on critical care capacity over the next 5 years.

\section{Conclusion}
In this paper, we leverage the huge amount of human mobility data collected from a widely-used Web mapping platform in China, to look into the significant change of people's transportation-related behaviors during the COVID-19 pandemic. To this end, we studied and performed data analysis on extensive transportation-related behaviors including the means of transportation, the type of visited venues, the check-in time of the venues, the preference on ``origin-destination'' distance, and ``origin-transportation-destination'' patterns. The data-driven analysis in this paper was conducted by comparing how those factors performed in the years of 2018, 2019, and 2020. Generally speaking, without the impact of the COVID-19 pandemic, the data distributions of these factors are consistent in 2018 and 2019. However, during the period of the COVID-19 pandemic in Mainland China (i.e., from January 2020 to April 2020), the data distributions of these factors dramatically change compared with those in the year 2019, indicating that the COVID-19 pandemic did cause great impact on the transportation-related behaviors of the public in Mainland China. For each factor, we also list our data-driven insights and policy-making suggestions on fighting this disease as follows:
\begin{itemize}
    \item \textit{Means of transportation}: Given the possibility of future outbreaks of COVID-19 or other emerging infectious diseases, the authorities could encourage more people to travel by bicycle and consider temporarily expanding the dedicated bicycle lanes. Moreover, they may coordinate the working hours of various companies, and suggest the public to travel in different peaks.
    \item \textit{Type of visited venues}: For residential areas as well as hospitals \& pharmacies, the authorities need to invest more effort in these venues to limit transmission to susceptible individuals and minimize the risk of cross-infection. After the pandemic is over, more economic support policies should be formulated for restaurants \& beverages, parks \& scenic spots, educational institutes, and hotels.
    \item \textit{Check-in time of venues}: As the people in Mainland China prefer going outdoors in the days of Wednesday and Thursday during the pandemic rather than in the days of weekends, the authorities should pay more attention to the daylight hours on Wednesday and Thursday rather than the daylight hours of weekends.
    \item \textit{Preference on ``origin-destination'' distance}: Due to the panic of the public on the pandemic, the proportion of OD distance $>300km$ in February 2020 is significantly greater than those in February 2018 and February 2019. The containment policy imposed by the Chinese authorities has been proved to effectively control the spread of COVID-19, which is also evidenced by the declined proportion of inter-city navigation requests in March and April in 2020.
    \item \textit{``Origin-transportation-destination'' patterns}: The start and end points of navigation by walk and private vehicle, are replaced by residential areas from transport facilities. This is the most dramatic change in people's travel patterns (i.e., the top-1 OTD pattern) during the pandemic. It indicates that people prefer zero-touch transportation to and from their places of residence. The changes of top-5 OTD patterns also tell us that the COVID-19 pandemic causes markets and workplaces to be the preferred destinations rather than hotels and educational institutes when people decide to go outside. We consider this phenomenon may be partly due to the containment policy, which stimulates the need to stockpile daily necessities and resume work. Therefore, there is a strong need to invest more effort in markets and workplaces. 
    The increased proportion of taking public transit to hospitals \& pharmacies should draw much attention from the authorities, as this OTD pattern tends to exacerbate the risk of infectious disease transmission. Hence, we propose to strengthen the prevention measures at transport facilities. 
\end{itemize}

\section{Future Work}
The navigation record is the major data resource employed by this paper to quantify the impact of COVID-19 on transportation-related behaviors of the public. It has helped us come up with many well-directed insights and policy-making suggestions on fighting this pandemic.
Besides the navigation records~\citep{smith2003web}, human mobility data also include the POI search logs~\citep{10.1145/3394137,huang-pac-2020}, rise and fall of POIs~\citep{fan2019monopoly,lu2016characterizing}, etc. In the future, we believe that each of those aspects of human mobility data can help us better understand other behaviors of the public during the period of the COVID-19 pandemic.
Furthermore, human mobility data can provide real-time evidence on the real-world behaviors of the public. This also inspires us to build an AI assistant for epidemic control and policy-making suggestion. 

\balance
\bibliographystyle{ACM-Reference-Format}
\bibliography{main}


\end{document}